\documentclass{iau}
\usepackage{graphicx}

\title[Stellar opacities] 
{Testing microphysics data}

\author[Walczak \& Daszy\'nska-Daszkiewicz]   
{Przemys{\l}aw Walczak
 \and Jadwiga Daszy\'nska-Daszkiewicz}

\affiliation{Instytut Astronomiczny, Uniwersytet Wroc{\l}awski, ul. Kopernika 11, 51-622 Wroc{\l}aw, Poland \\ email: {\tt walczak@astro.uni.wroc.pl},
{\tt daszynska@astro.uni.wroc.pl}}

\pubyear{2013}
\volume{301}  
\pagerange{1--8}
\jname{Precision Asteroseismology}
\editors{J.A. Guzik, W.J. Chaplin, G. Handler \& A. Pigulski, eds.}
\begin{document}

\maketitle

\begin{abstract}
High precision asteroseismic data provide a unique opportunity to test input microphysics such as stellar opacities, chemical composition or equation of state. These tests are possible because pulsational frequencies as well as amplitudes and phases of the light variations are very sensitive to the internal structure of a star. We can therefore compute pulsation models and compare them with observations. The agreement or differences should tell us whether some models are adequate or not, and which input data need to be improved.
\end{abstract}

\firstsection 
\section{Introduction}


One of the most important ingredients of the stellar input physics are opacities. The value of the opacity coefficient affects the pulsational properties of models. There are two basic asteroseismic tools: frequencies and the corresponding values of the complex non-adiabatic $f$-parameter. The $f$-parameter if defined as the ratio of the radiative flux change to the radial displacement at the photospheric level \cite[(Daszy\'nska-Dasziewicz et al.~2003, 2005)]{DD03,DD05}. Its value determines amplitudes and phases of light variations. In case of B-type stars, the empirical value of the
$f$-parameter can be determined only for modes that are visible both in multicolor photometry and spectroscopy. It is important to add that the empirical $f$-parameter depends slightly on the input from model atmospheres. The discussion of these effects can be found in \cite{JDDWS2012} and \cite{JDD2013}. Here, in all computations we adopt the LTE models of stellar atmospheres by \cite{K04} with microturbulence velocity of $\xi_t=2$ km\,s$^{-1}$.

\begin{figure}[h]
 \vspace*{-0.2 cm}
\begin{center}
\includegraphics[clip,width=63mm]{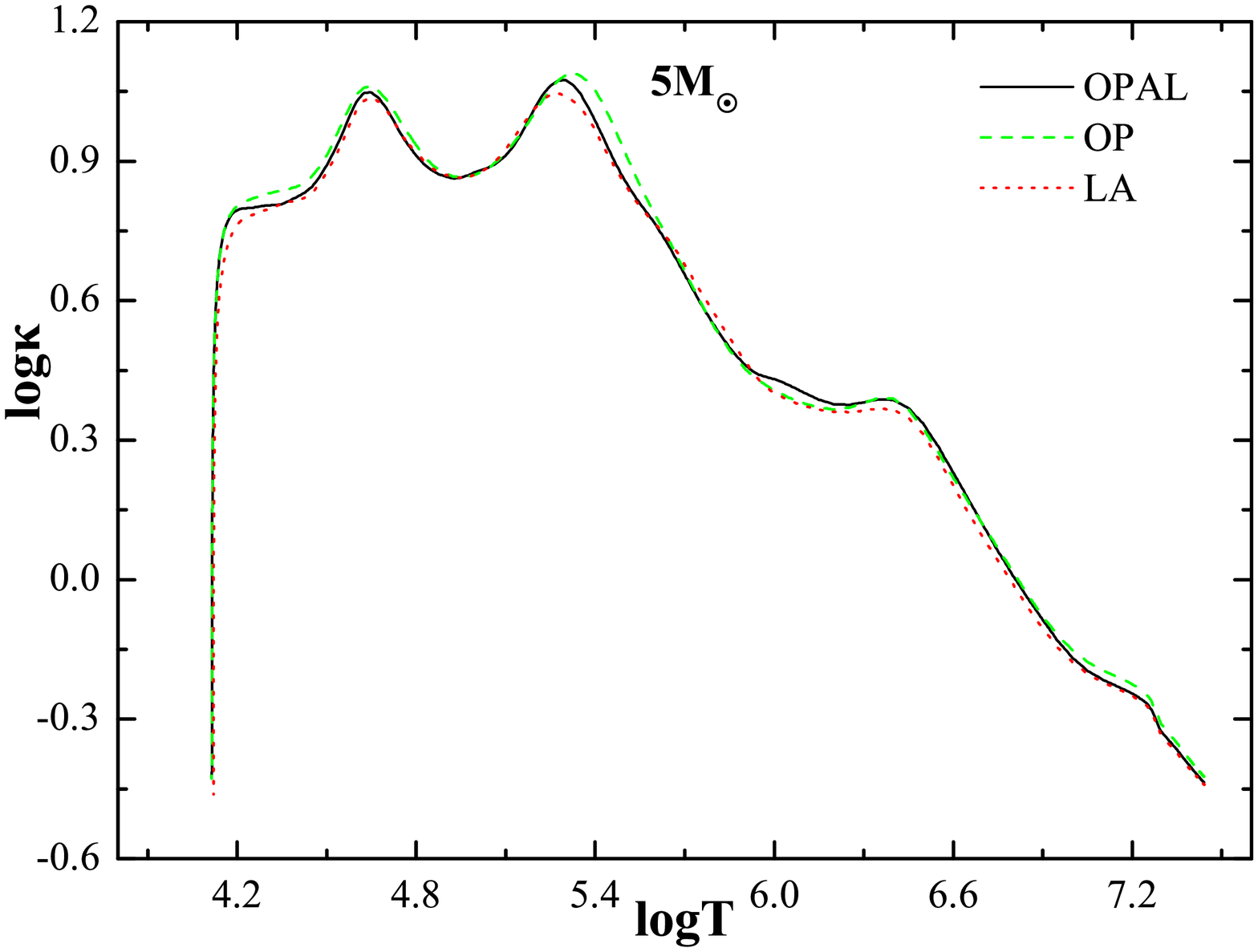}
 \includegraphics[clip,width=63mm]{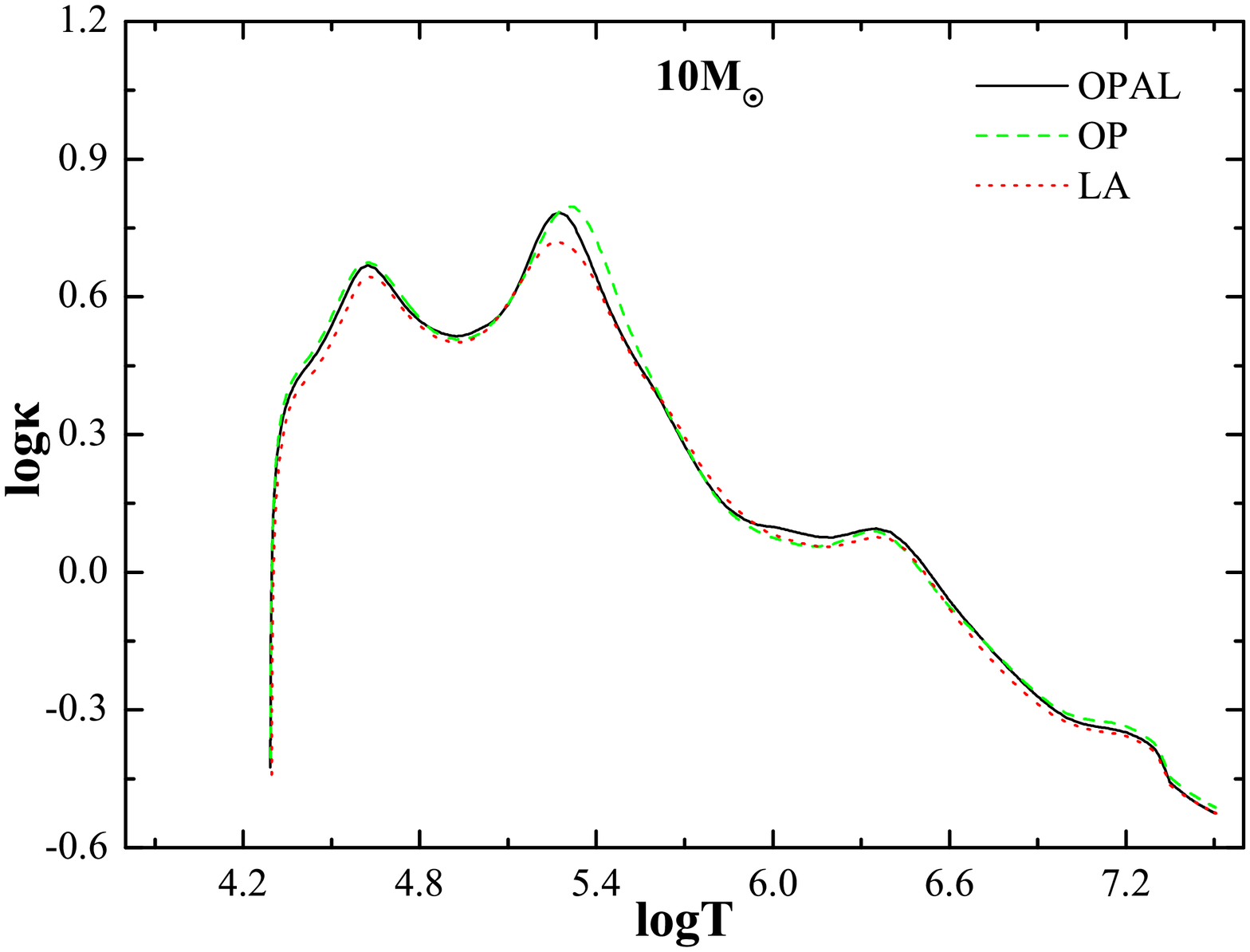}
  \vspace*{-0.2 cm}
\caption{The Rosseland mean opacity, $\kappa$, as a function of the temperature, $T$, inside of the 5\,$M_{\odot}$ (left panel) and 10\,$M_{\odot}$ (right panel) stellar models with effective temperatures $\log{T_{\rm{eff}}}\sim4.196$ and $\log{T_{\rm{eff}}}\sim4.373$, respectively. Three sources of the opacity data were considered: the OP, OPAL and LA tables.}
   \label{opacity}
\end{center}
\end{figure}

We tested three available opacity tables: OP \cite[(Seaton 2005)]{OP}, OPAL \cite[(Iglesias \& Ro\-gers 1996)]{OPAL} and new data from Los Alamos (LA) \cite[(Magee et al.~1995)]{LA}. In Fig.\,\ref{opacity}, we show a comparison of the Rosseland mean opacity, $\kappa$, plotted as a function of the temperature within stellar models. The metallicity parameter was set to $Z=0.02$. In the left and right panels we considered stellar models of $5M_{\odot}$, $\log{T_{\rm{eff}}}\sim4.196$ and $10M_{\odot}$, $\log{T_{\rm{eff}}}\sim4.373$, respectively. 


We can easily notice two high opacity bumps. The first one, occurring at the lower temperature, is connected with ionization of HeII.
The second one,  the so-called $Z$ bump, is caused by a large number of transition lines of iron-group elements. Although
there are some small differences near the $Z$ bump, the OP and OPAL data are quite similar. On the other hand,
the LA opacity coefficient is in general smaller than the OP and OPAL ones, especially in the region of the $Z$ bump.
As one can expect, this fact has huge consequences on pulsational instability in B-type stars.

\section{Inferring constraints on opacities}

\begin{figure}[h]
 \vspace*{-0.2 cm}
\begin{center}
\includegraphics[clip,width=103mm]{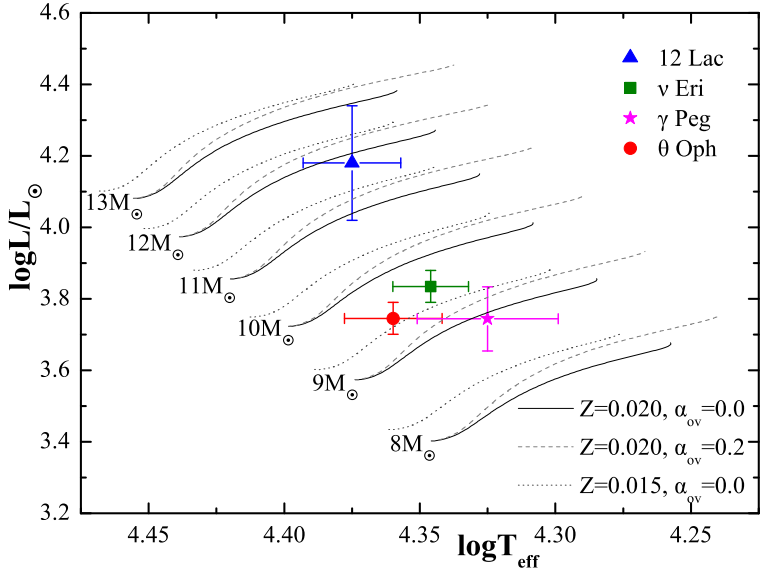}
 \vspace*{-0.2 cm}
 \caption{The H-R diagram with positions of four $\beta$ Cephei stars: $\theta$ Oph, $\gamma$ Peg, 12 Lac and $\nu$~Eri. The theoretical evolutionary tracks for masses from $8$ to 13\,$M_{\odot}$ were calculated for different values of metallicity, $Z$, and overshooting parameter, $\alpha_{\rm{ov}}$. Only the main sequence part of evolution is shown.}
   \label{fig1}
\end{center}
\end{figure}

We have chosen four $\beta$ Cephei-type stars for our tests: $\theta$ Ophiuchi (HD\,157056), $\gamma$~Pegasi (HD\,886), 12 Lacertae (HD\,214993) and $\nu$ Eridani (HD\,29248). In Fig.\,\ref{fig1}, we plot their positions  in the Hertzsprung-Russell diagram. We added also evolutionary tracks from ZAMS to TAMS for masses from $8$ up to $13M_{\odot}$ with step $1M_{\odot}$. The tracks were calculated with the OP opacity tables, two values of metallicity ($Z=0.015$ and $0.020$) and two values of the overshooting parameter ($\alpha_{\rm{ov}}=0.0$ and $0.2$). 
Unless otherwise noted, in all computations we assumed the chemical composition by \cite{AGSS09} and the initial hydrogen abundance $X=0.7$.

In Fig.\,\ref{fig1} we can easily see that the masses of the stars are from about $9M_{\odot}$ to about $12M_{\odot}$. All stars are most likely in the core hydrogen-burning evolution phase. We can also notice the huge impact of metallicity and the overshooting parameter on the theoretical tracks. The lower the metallicity, the higher the mass that can be derived from the H-R diagram. A high value of the overshooting parameter prolongs the duration of the main-sequence phase.

\subsection{$\theta$ Ophiuchi}

$\theta$ Oph is a $\beta$ Cephei-type star that pulsates in at least seven frequencies \cite[(Handler et al.~2005)]{HSM05}. Three of them were also found in the radial velocity measurements \cite[(Briquet et al.~2005)]{B05}. The effective temperature of $\theta$ Oph, $\log{T_{\rm{eff}}}=4.360\pm0.018$, was determined by \cite{HSM05}. The luminosity, $\log{L/L_{\odot}}=3.746\pm0.045$, was calculated taking into account the \textit{Hipparcos} parallax $\pi=7.48\pm0.17$ mas and the bolometric correction from the calibration by \cite{Flower1996}.

\begin{figure}[h]
 \vspace*{-0.2 cm}
\begin{center}
\includegraphics[clip,width=67mm]{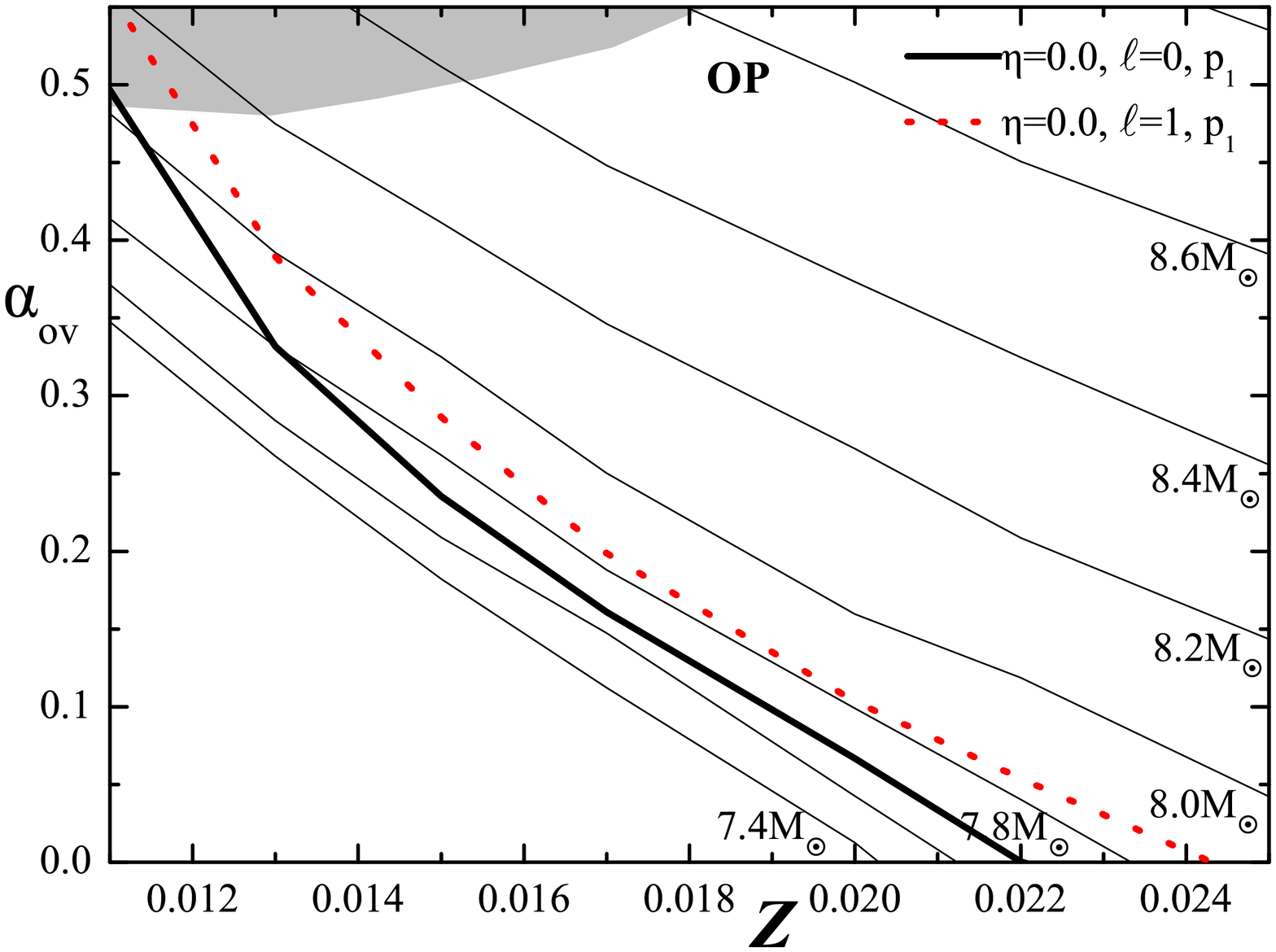}
\includegraphics[clip,width=67mm]{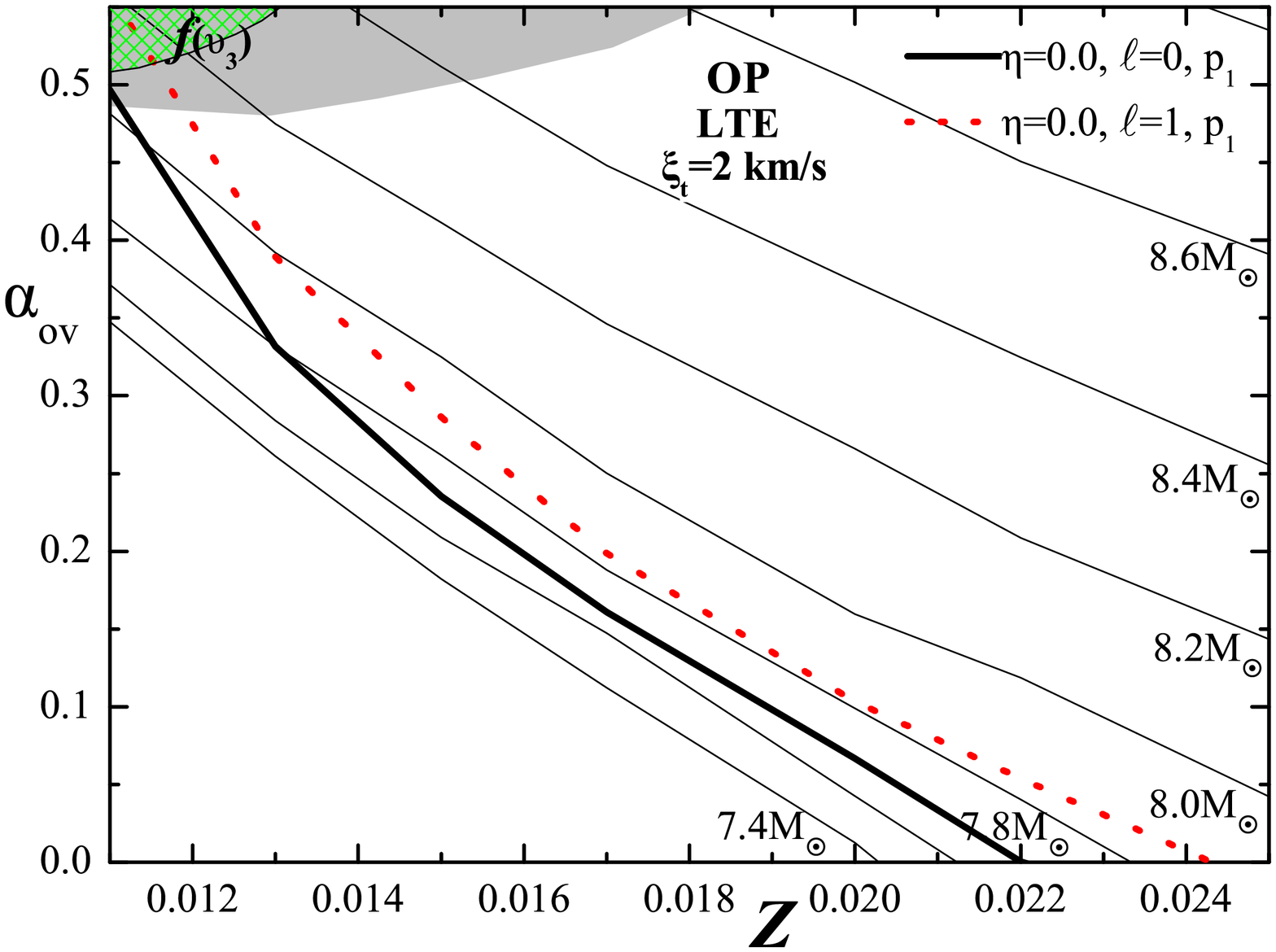}
 \includegraphics[clip,width=67mm]{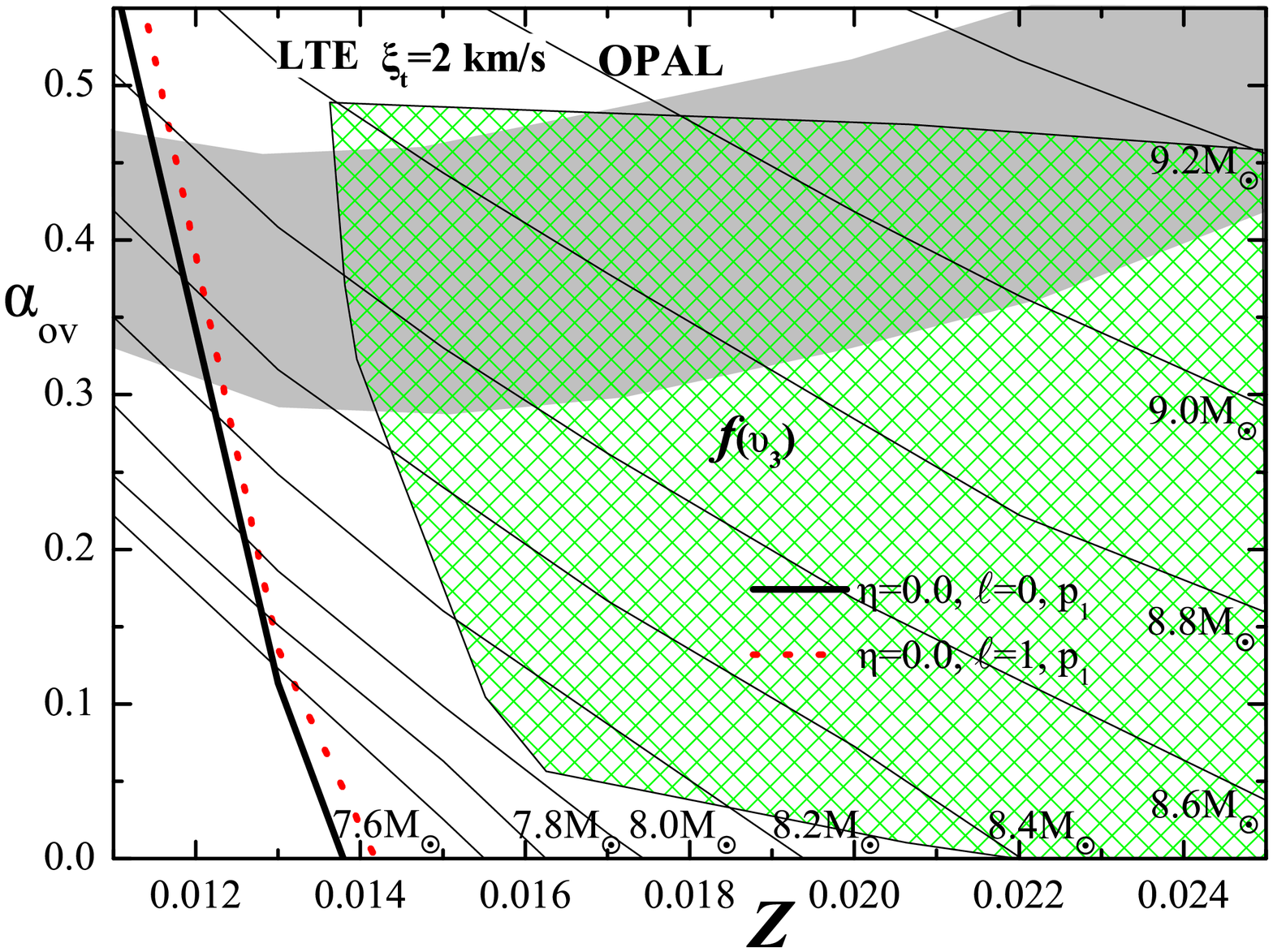}
\includegraphics[clip,width=67mm]{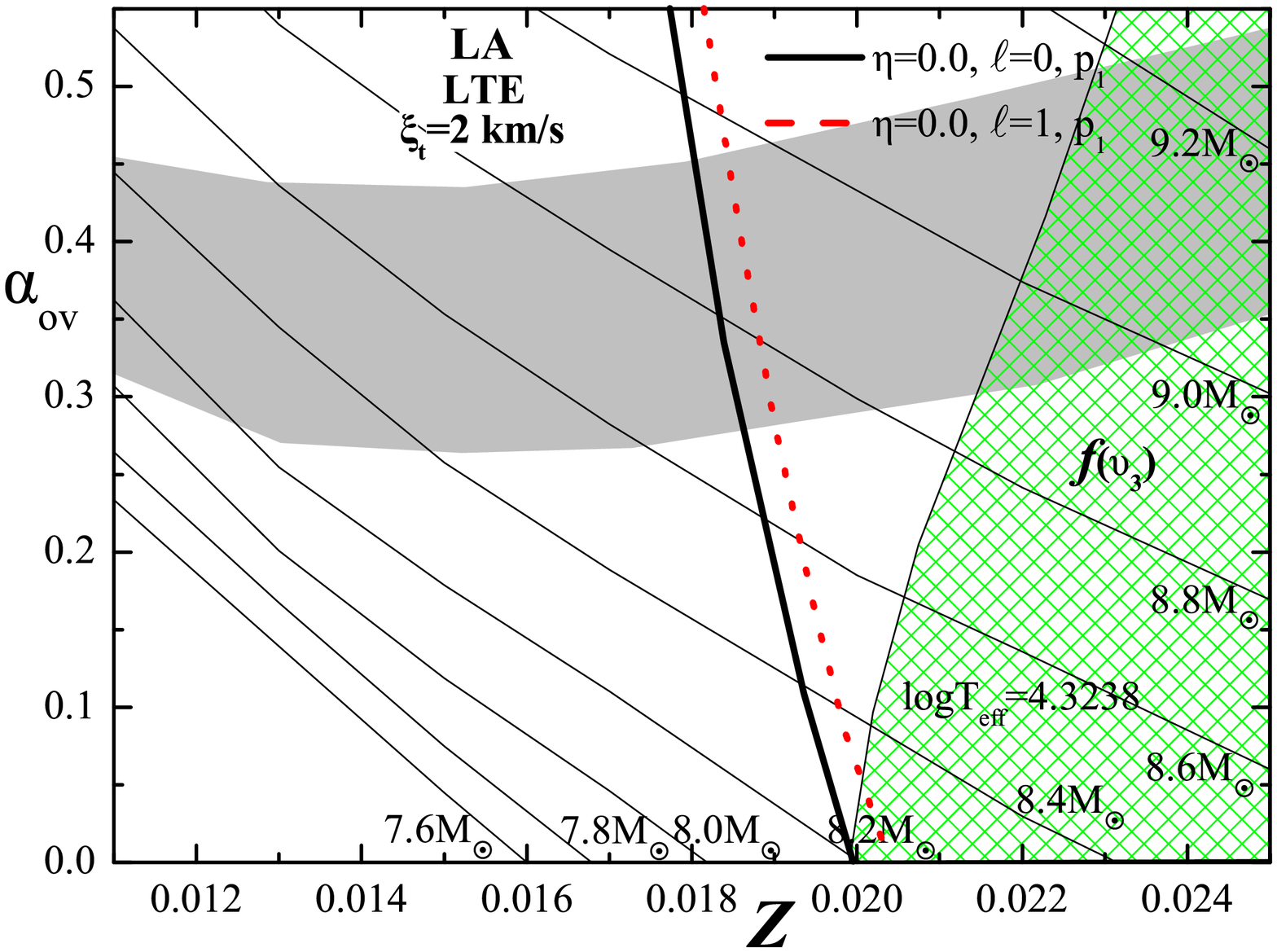}
 \caption{Seismic models of $\theta$ Oph fitting two frequencies ($\nu_3=7.4677$ d$^{-1}$ as an $\ell=0$ p$_1$ mode and $\nu_6=7.8742$ d$^{-1}$ as $\ell=1$, p$_1$ mode) on the overshooting ($\alpha_{\rm{ov}}$) $vs.$ metallicity ($Z$) planes. Grey areas indicates models lying inside of the observational error box. We show also lines of constant mass (thin solid lines) and instability borders for the radial mode (thick solid line) and dipole mode (thick dashed line). In the upper panels we used the OP opacity tables. In the upper right panel we marked models fitting additionally the empirical value of the $f$-parameter for $\nu_3$ (hatched area). The bottom panels are the same as the upper right panel, except that we used the OPAL (left panel) and LA (right panel) opacities.}
   \label{Ophfig1}
\end{center}
\end{figure}

In our modelling we used only axisymmetric modes (with azimuthal number $m=0$). In the case of $\theta$ Oph, mode identification indicates
two centroid modes: $\nu_3=7.4677$ d$^{-1}$ (radial p$_1$) and $\nu_6=7.8742$ d$^{-1}$ (dipole p$_1$) \cite[(Daszy\'nska-Daszkiewicz \& Walczak 2009)]{DW09}.
We have found models fitting  these two frequencies for different values of metallicity, $Z$, and the overshooting parameter, $\alpha_{\rm{ov}}$.
The results for the OP opacity tables are shown in the upper left panel of Fig.\,\ref{Ophfig1} on the overshooting
parameter ($\alpha_{\rm{ov}}$) $vs.$ metallicity ($Z$) plane. We marked lines of constant masses (thin lines) and instability borders
for the radial (thick solid line) and dipole (thick dashed line) modes. Models that lay above these instability borders are excited.
Instability borders were defined as the zero value of the instability parameter, $\eta={W}/{\int_0^R \left|\frac{dW}{dr}\right|dr}$, where $W$ is the work integral and $R$ is the stellar radius. It can be seen that there exist a lot of models fitting the $\nu_3$ and $\nu_6$ frequencies. Only for the low values of metallicity and the overshooting parameter were we unable to find seismic models (bottom left corner of the panel). The grey area indicates models lying inside of the observational errors of the effective temperature and luminosity of $\theta$ Oph. Models below the grey area are cooler and less bright than the error box.


The radial mode, $\nu_3$, was detected both in photometry and spectroscopy. Therefore we were able to derive the empirical value of the non-adiabatic $f$-parameter for this mode. We compared this value with theoretical equivalent and found models fitting it (within the errors). In the upper right panel of Fig.\,\ref{Ophfig1} we showed the same figure as in the upper left panel, but in addition we marked models which also fit the empirical value of the $f$-parameter for the $\nu_3$ mode (hatched area labeled with $f(\nu_3)$).

As we can see, models fitting two frequencies ($\nu_3$ and $\nu_6$) and the $f$-parameter of the radial mode are located inside the observational error box. On the other hand, these models have a very efficient overshooting from the convective core, with $\alpha_{\rm{ov}}\sim0.5$, which is not expected in a rather slowly rotating star like $\theta$ Oph ($V_{\rm{rot}}\approx30$ km\,s$^{-1}$).

A different situation arises in the case of the OPAL and LA opacities (bottom left and right panel of Fig.\,\ref{Ophfig1}, respectively). For a given value of $Z$ and $\alpha_{\rm{ov}}$, OPAL and LA models have much higher masses than OP models. Also, the effective temperature and luminosity are larger, and models that are inside of the error box appear for less effective core overshoot ($\alpha_{\rm{ov}} \sim 0.3-0.5$).

With the OPAL opacities we were able to find quite a large number of models fitting the $f$-parameter of the radial mode $\nu_3$. Some of these models lie inside of the observational error box. For the case of the LA opacities, we managed to fit the $f$-parameter only for models with metallicity larger that about $0.02$. Moreover, for the LA opacity models with $Z$ lower than 0.02, the modes considered are stable.

In the case of $\theta$ Oph, the models turned out to be very sensitive to the differences between opacities. As we could see, the value of $\kappa$ has also a very large impact on the $f$-parameter.

\subsection{$\gamma$ Pegasi}

$\gamma$ Peg is a B2 spectral-type star that pulsates in at least 14 modes \cite[(Handler et al.~2009)]{handler2009}. Six of the modes have very low frequencies ($<0.9$ d$^{-1}$), typical for the Slowly Pulsating B-type stars. The remaining 8 modes are of the $\beta$ Cep-type. Because of this, the star is a hybrid pulsator of the $\beta$ Cep/SPB type.

The effective temperature ($\log T_{\rm eff}=4.325 \pm 0.026$) as well as the luminosity ($\log L/L_\odot$ $=3.744\pm0.09$) of $\gamma$~Peg, shown in Fig.\,\ref{fig1} were adopted from \cite{Walczak2013}. We chose two well identified $\beta$~Cep modes: the radial p$_1$ ($\nu_1=6.58974$ d$^{-1}$) and dipole g$_1$ ($\nu_5=6.01616$ d$^{-1}$), and constructed models fitting them.

\begin{figure}[h]
 \vspace*{-0.2 cm}
\begin{center}
\includegraphics[clip,width=67mm]{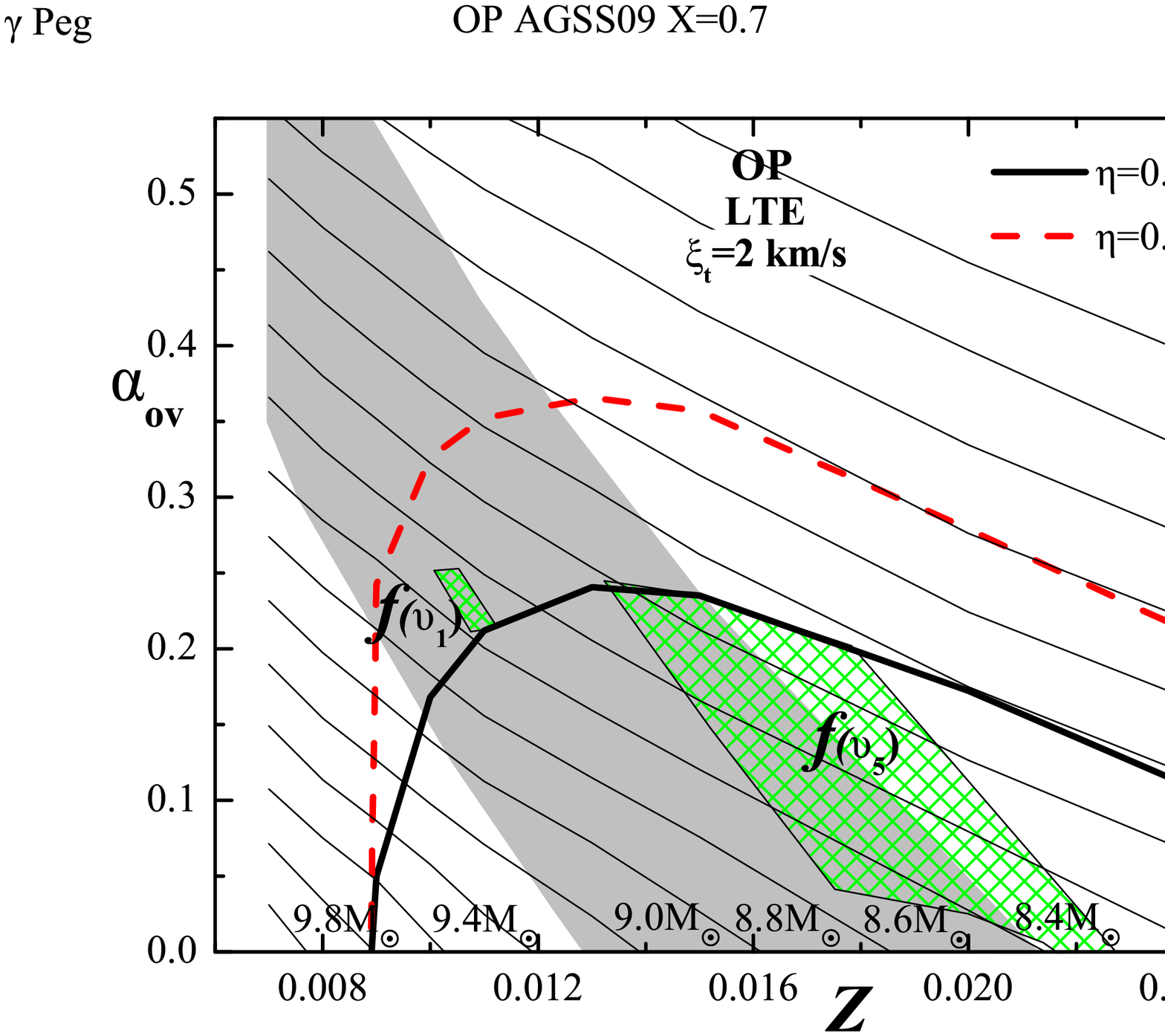}
 \includegraphics[clip,width=67mm]{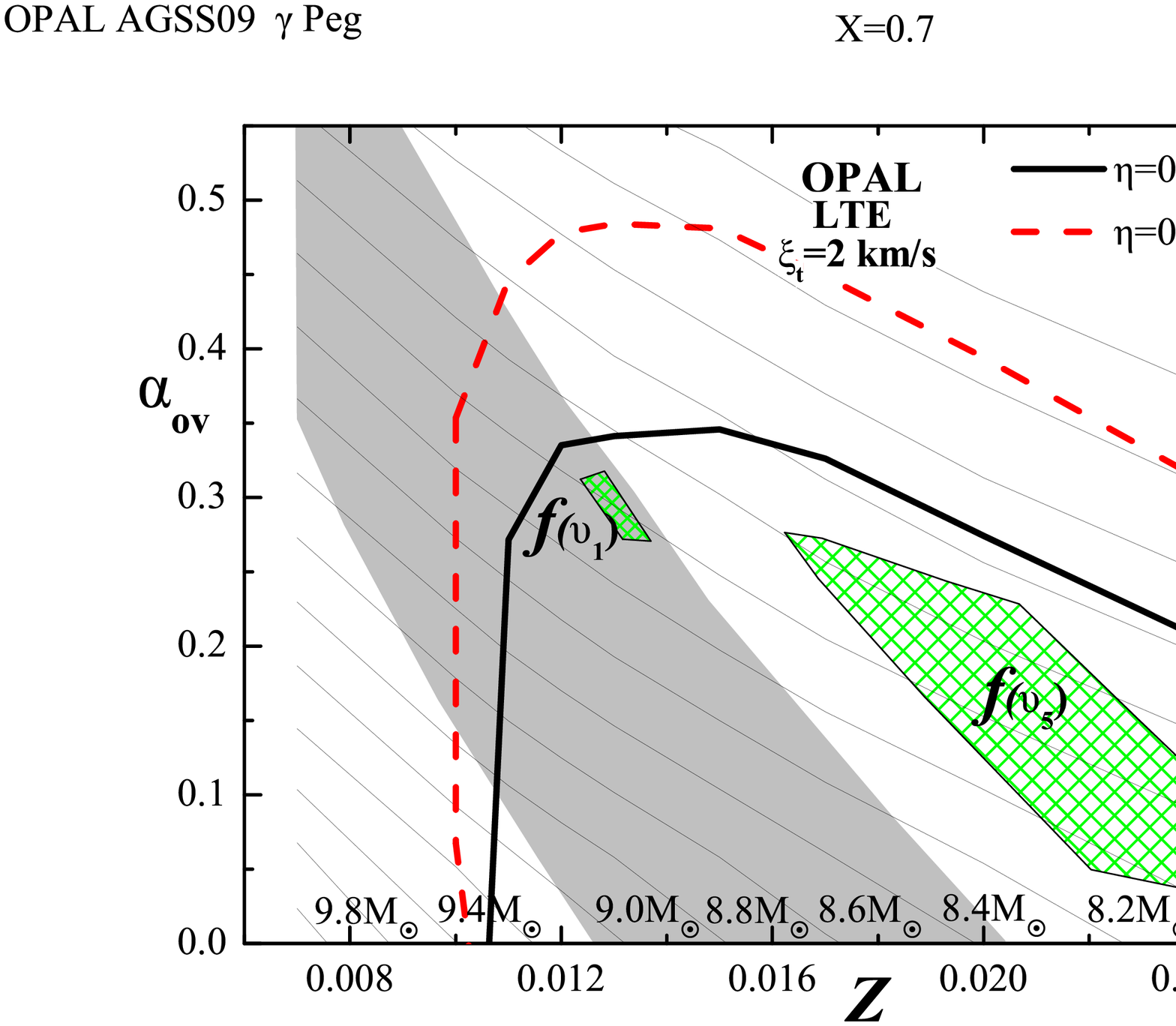}
 \includegraphics[clip,width=67mm]{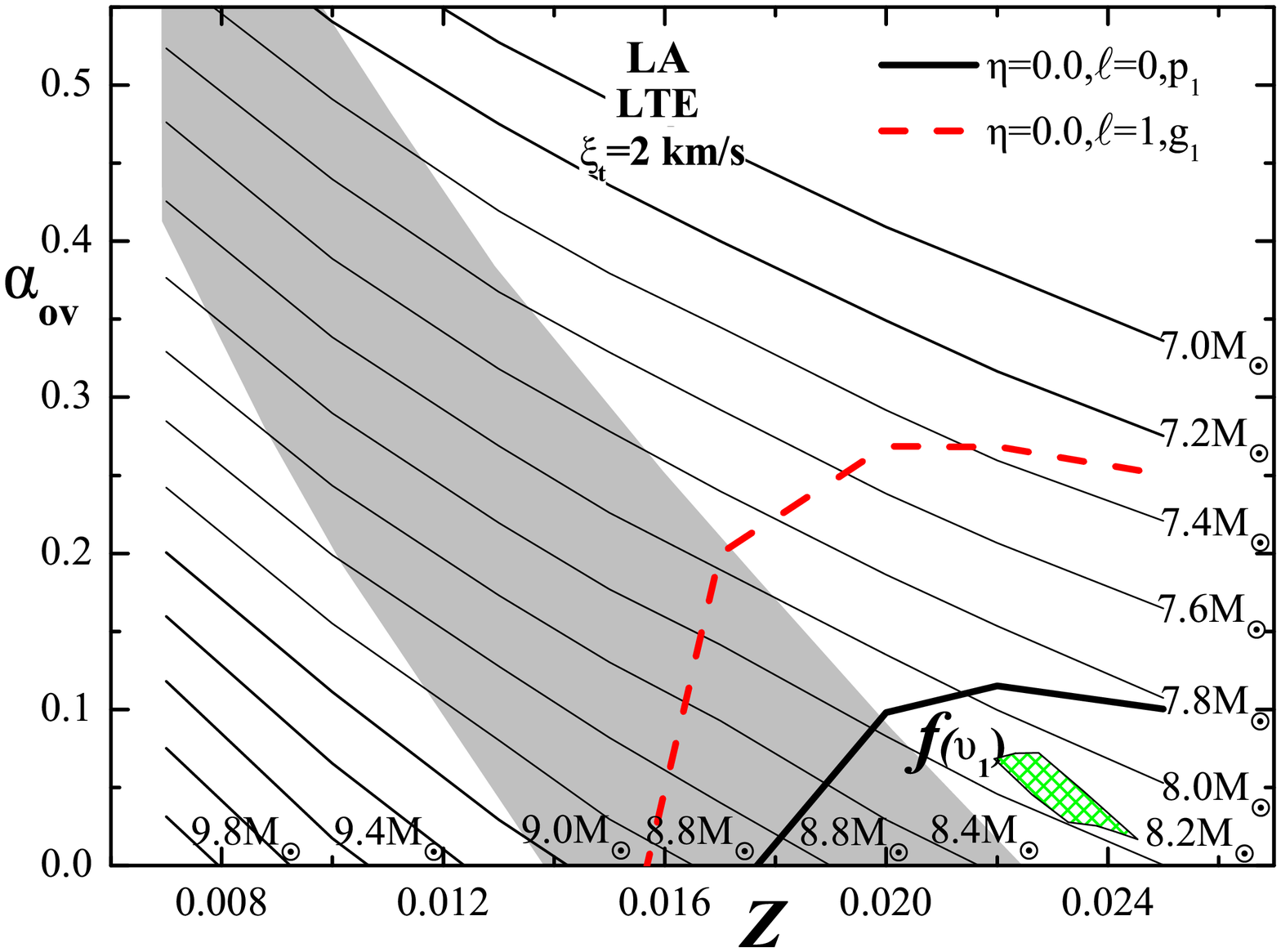}
 \vspace*{-0.2 cm}
 \caption{Seismic models of $\gamma$ Peg fitting two frequencies ($\nu_1=6.58974$ d$^{-1}$ as $\ell=0$ p$_1$ mode and $\nu_5=6.01616$ d$^{-1}$
 as $\ell=1$, g$_1$ mode) on the $\alpha_{\rm{ov}}$ $vs.$ $Z$ plane. In the upper left panel we used the OP opacities, in the upper right - OPAL
 and in the bottom - LA.} 
   \label{gPeg}
\end{center}
\end{figure}

The models are shown in Fig.\,\ref{gPeg} on the $\alpha_{\rm{ov}}$ $vs.$ $Z$ plane in three panels corresponding to computations
with the OP, OPAL and LA opacities. Here, the unstable modes are below the drawn instability borders. For $\gamma$ Peg we were also able to determine the empirical values of the $f$-parameter for the $\nu_1$ and $\nu_5$ modes. They are marked in Fig.\,\ref{gPeg} as hatched areas labeled as $f(\nu_1)$ and $f(\nu_5)$. Unfortunately, we could not find a single model that would fit the empirical values of the $f$-parameter for these two modes simultaneously. The OP and OPAL models are quite similar. There is only a difference in position of models fitting the $f$-parameter. In the case of the OPAL data the hatched areas are shifted to a higher value of metallicity. With the LA data, however, we did not find models fitting the empirical value of the $f$-parameter for the $\nu_5$ mode (in the metallicity range considered, $Z\in 0.007-0.025$). The LA models fitting the $f$-parameter for $\nu_1$ mode have high metallicities, $Z\approx 0.023$,  and are outside of the observational error box.

\subsection{12 Lacertae}

12 Lac is a well known pulsating $\beta$ Cep/SPB type star. It pulsates in 11 modes \cite[(Handler et al.~2006)]{handler2006}. One mode is SPB-type. In Fig.\,\ref{fig1}, we showed the error box of 12 Lac. The value of the effective temperature ($\log{T_{\rm{eff}}}=4.375 \pm0.018$) and luminosity ($\log L/L_\odot=4.18 \pm 0.16$) was taken from \cite{handler2006}.

\begin{figure}[h]
 \vspace*{-0.2 cm}
\begin{center}
\includegraphics[clip,width=67mm]{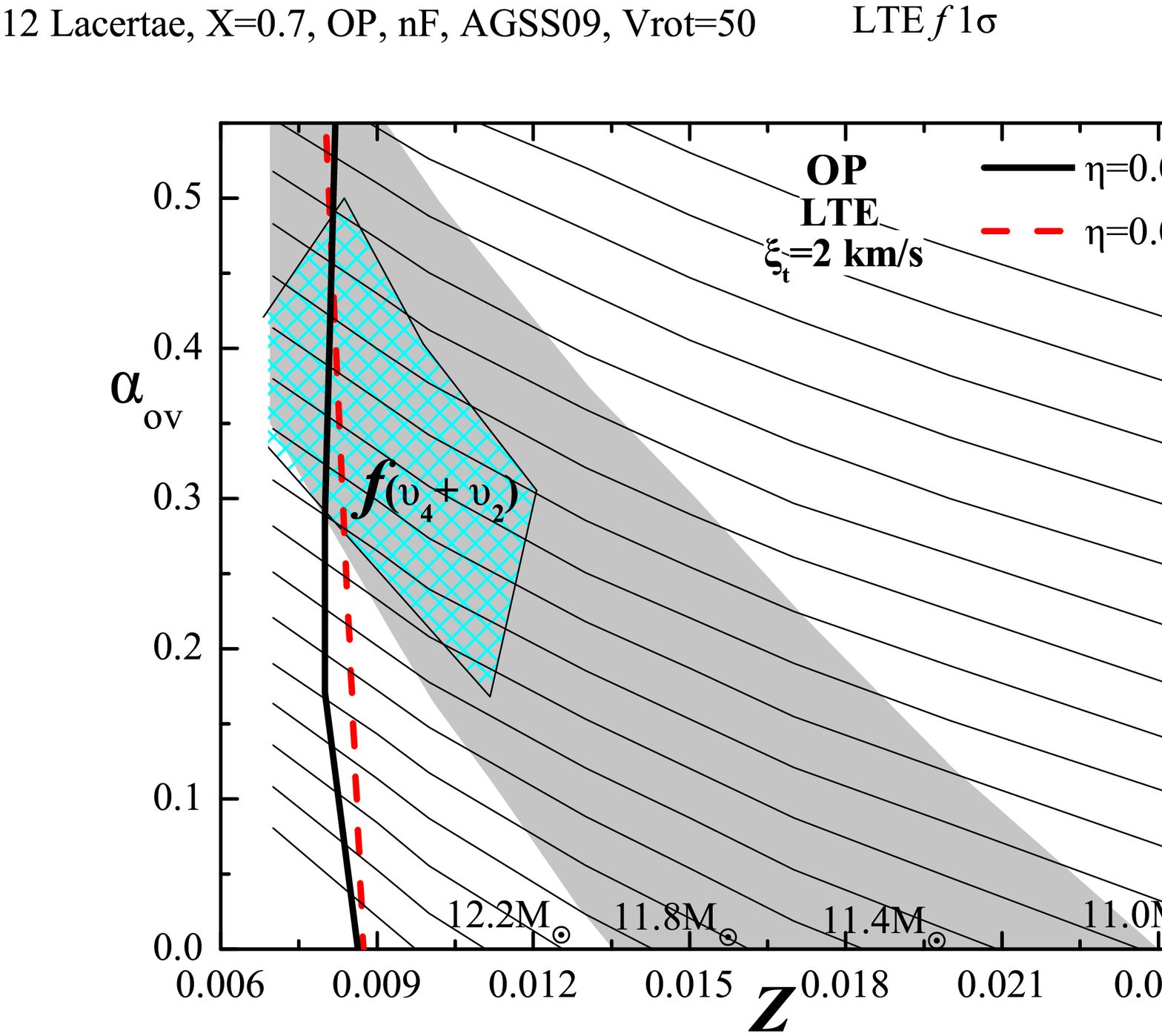}
 \includegraphics[clip,width=67mm]{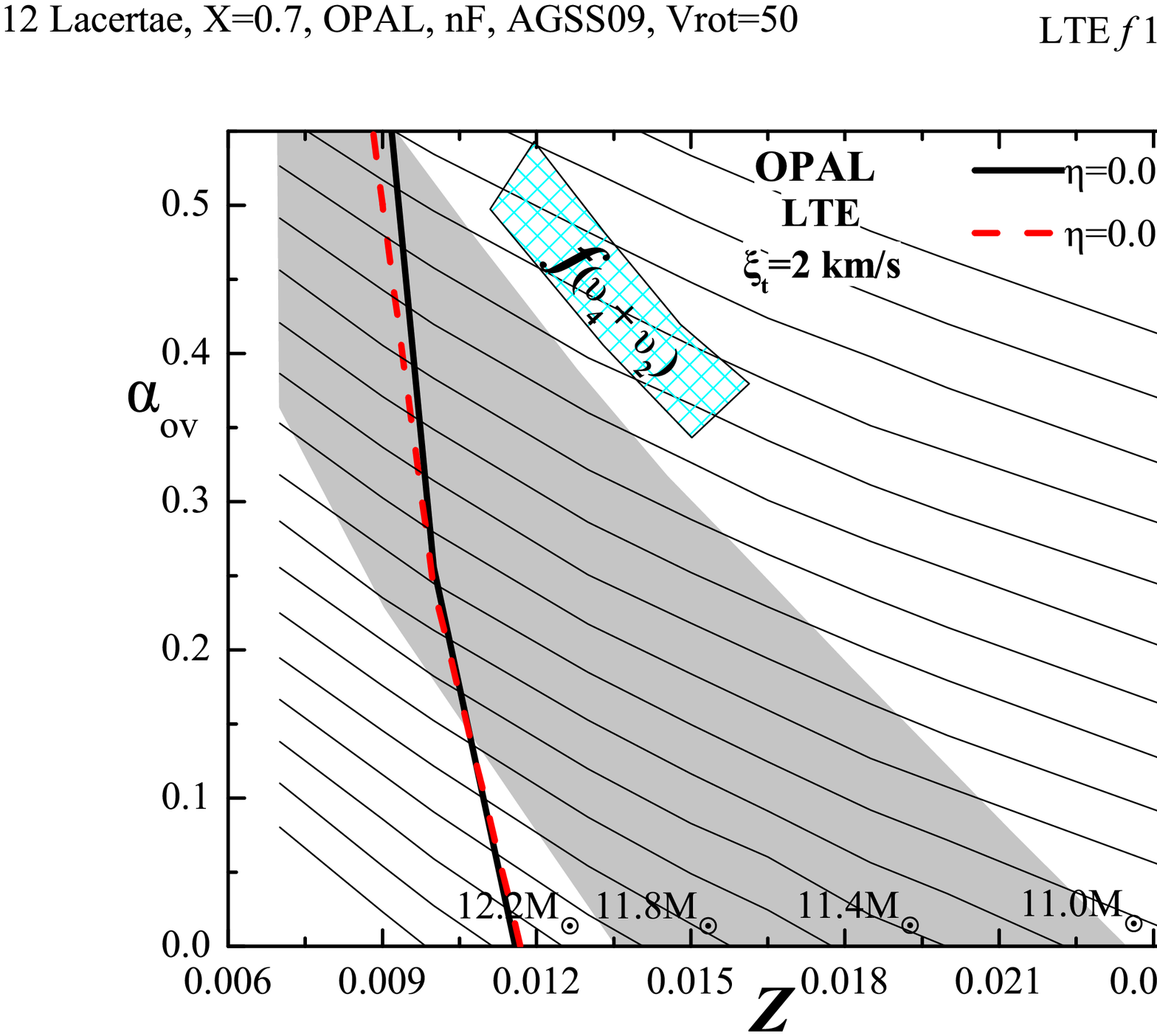}
 \includegraphics[clip,width=67mm]{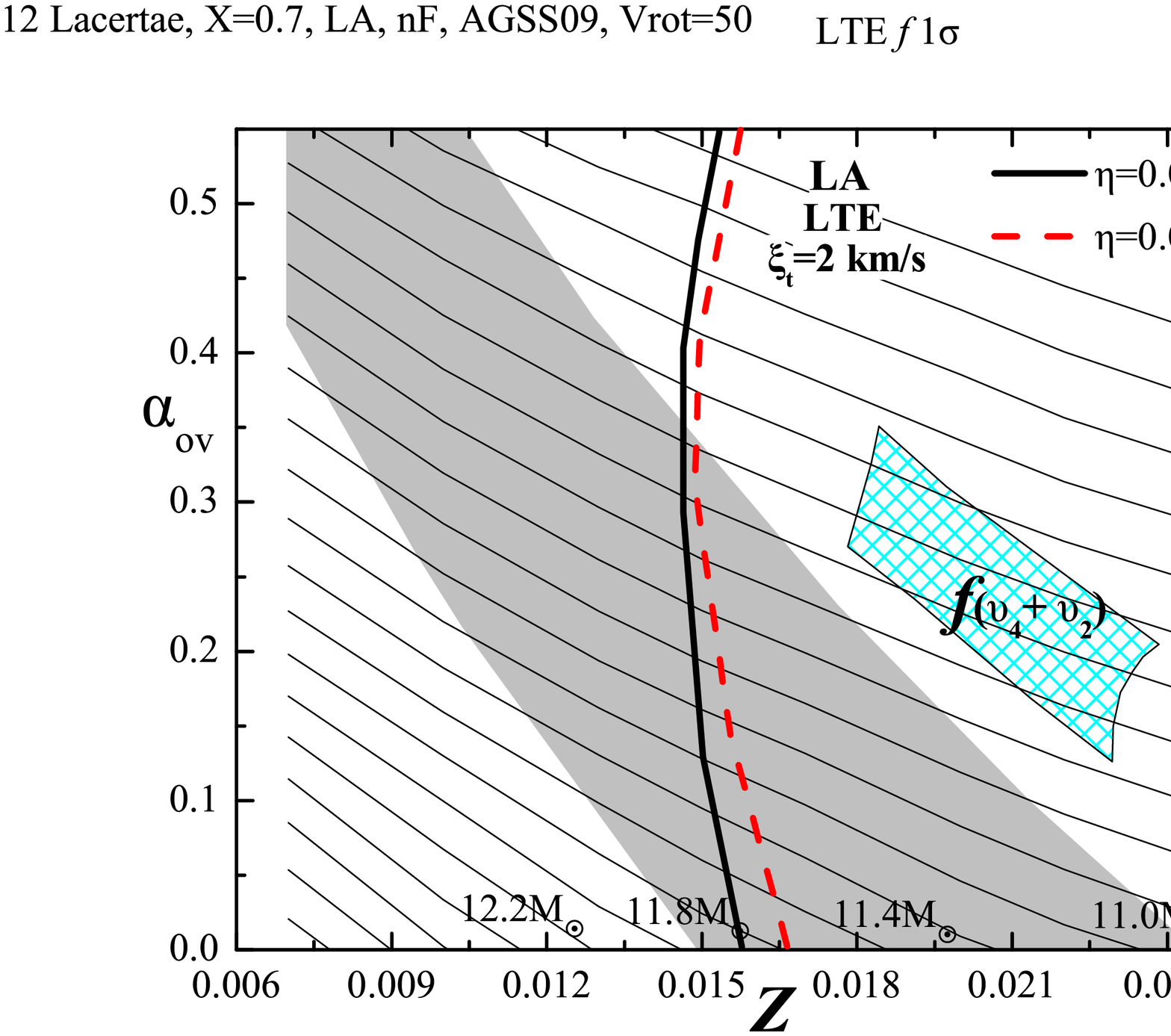}
 \vspace*{-0.2 cm}
 \caption{Seismic models of 12 Lac fitting two frequencies ($\nu_2=5.066346$ d$^{-1}$ as $\ell=1$ g$_1$ mode and $\nu_4=5.334357$ d$^{-1}$ as $\ell=0$, p$_1$ mode) on the $\alpha_{\rm{ov}}$ $vs.$ $Z$ plane. In the upper left panel we used the OP opacities, in the upper right - OPAL and in the bottom - LA.}
   \label{12Lac}
\end{center}
\end{figure}

Based on the mode identification \cite[(Daszy\'nska-Daszkiewicz et al.~2013)]{JDD2013}, we know that at least two modes are axisymmetric: $\nu_2=5.066346$ d$^{-1}$ ($\ell=1$, g$_1$) and $\nu_4=5.334357$ d$^{-1}$ ($\ell=0$, p$_1$). Models fitting these two frequencies are plotted in Fig.\,\ref{12Lac} on the $\alpha_{\rm{ov}}$ $vs.$ $Z$ plane. We managed to derived the value of the empirical $f$-parameter for these two modes. We were also able to find models which fit the $f$-parameter for these two modes simultaneously (hatched regions in Fig.\,\ref{12Lac}).

As we can see, in the case of the OP data (upper left panel of Fig.\,\ref{12Lac}), there are plenty of models fitting two frequencies ($\nu_2$ and $\nu_4$) and their $f$-parameters that are inside of the error box (grey area). The OPAL models fitting additionally the $f$-parameter for $\nu_2$ and $\nu_4$ are outside of the error box (upper right panel of Fig.\,\ref{12Lac}). A similar situation occurred in the case of the LA opacities. Models fitting the $f$-parameters are outside of the error box.

\subsection{$\nu$ Eridani}

$\nu$ Eri is one of the most extensively studied $\beta$ Cep/SPB pulsators. As is the case $\gamma$ Peg, this star pulsates in at least 14 modes \cite[(Handler et al.~2004; Jerzykiewicz et al.~2005)]{handler2004,J05}. Two of them are SPB-type. This star pulsates in three well identified centroid modes \cite[(e.g., Daszy\'nska-Daszkiewicz \& Walczak 2010)]{JDDPW2010}: one radial p$_1$ mode ($\nu_1=5.7632828$ d$^{-1}$), and two dipoles g$_1$ ($\nu_4=5.6372470$ d$^{-1}$) and p$_1$ ($\nu_6=6.243847$ d$^{-1}$).
There is also a $\nu_9=7.91383$ d$^{-1}$ mode, which could be a centroid of the dipole p$_2$ mode.

The effective temperature of $\nu$ Eri, $\log{T_{\rm{eff}}}=4.346\pm0.014$, was adopted from \cite{DD05}. The luminosity, $\log{L/L_{\odot}}= 3.835\pm0.045$, was calculated with the \textit{Hipparcos} parallax $\pi=483\pm19$ mas \cite[(van Leeuwen 2007)]{vanLeeuwen2007}. We used also the \cite{Flower1996} bolometric correction corresponding to the effective temperature of $\nu$ Eri.

\begin{figure}[h]
 \vspace*{-0.2 cm}
\begin{center}
\includegraphics[clip,width=66mm]{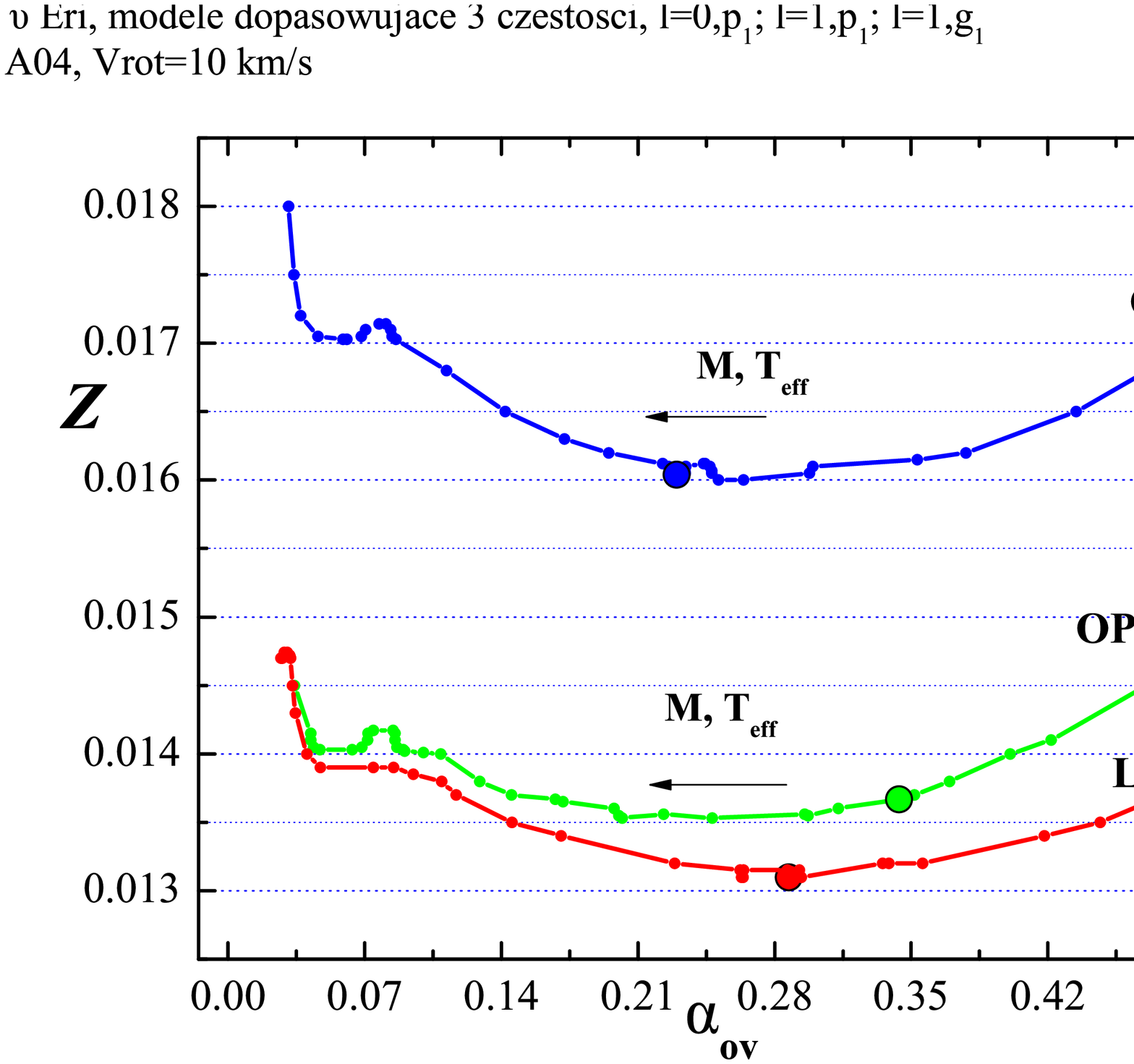}
 \includegraphics[clip,width=66mm]{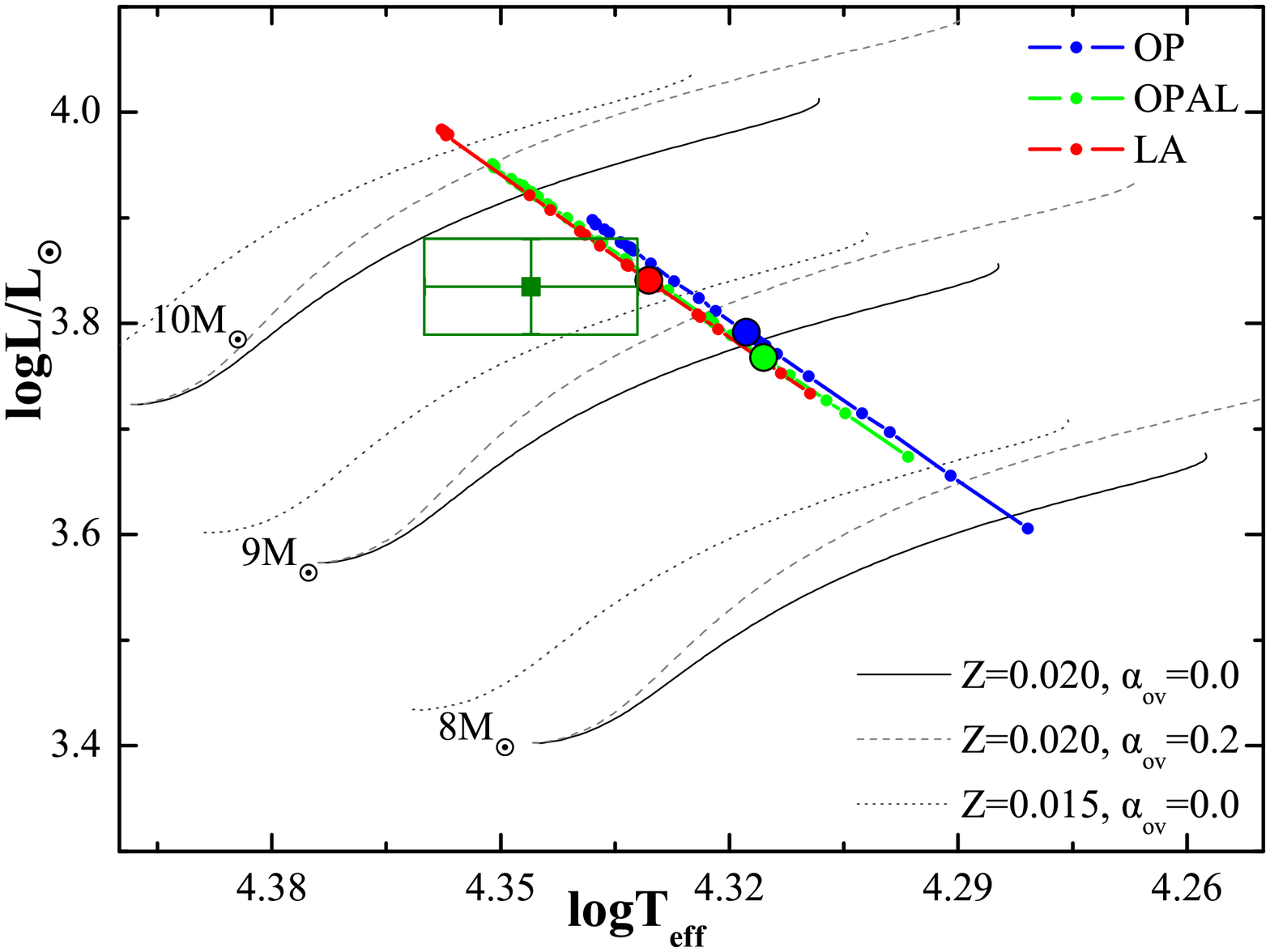}
 \vspace*{-0.2 cm}
 \caption{Left panel: seismic models of $\nu$ Eri fitting three frequencies ($\nu_1=5.7632828$ d$^{-1}$ as $\ell=0$ p$_1$ mode, $\nu_4=5.6372470$ d$^{-1}$ as $\ell=1$ g$_1$ mode and $\nu_6=6.243847$ d$^{-1}$ as $\ell=1$, p$_1$ mode) on the $Z$ $vs.$ $\alpha_{\rm{ov}}$ plane. Right panel: H-R diagram with position of seismic models of $\nu$ Eri.}
   \label{nuEri1}
\end{center}
\end{figure}

In the left panel of Fig.\,\ref{nuEri1}, we plotted models fitting three frequencies of $\nu$ Eri ($\nu_1$, $\nu_4$ and $\nu_6$).
The results are presented on the $Z$ $vs.$ $\alpha_{\rm{ov}}$ plane.
The large dots correspond to models fitting additionally also the $\nu_9$ frequency. Because we used three frequencies, we have lines of models instead of a plane, like in the case of $\theta$ Oph, $\gamma$ Peg or 12 Lac. Four frequencies reduce the number of models to one point in this kind of figure. We marked also the direction of increasing mass and effective temperature.

We can easily notice that seismic models calculated with different opacity tables are well separated in metallicity. The highest values of $Z$ occur for the OP data ($Z=0.016-0.018$). With OPAL opacities the metallicity is in the range from 0.0135 to 0.0145. The lowest metallicity was found with the LA tables: $Z=0.013-0.0145$. In the right panel of Fig.\,\ref{nuEri1}, we marked seismic models of $\nu$~Eri on the H-R diagram. We showed also the $\nu$~Eri error box. Unfortunately, seismic models are only partially inside the error box. It is interesting that the LA models have somewhat higher values of the effective temperature and fit the observational parameters slightly better.


\begin{figure}[h]
 \vspace*{-0.2 cm}
\begin{center}
\includegraphics[clip,width=63mm]{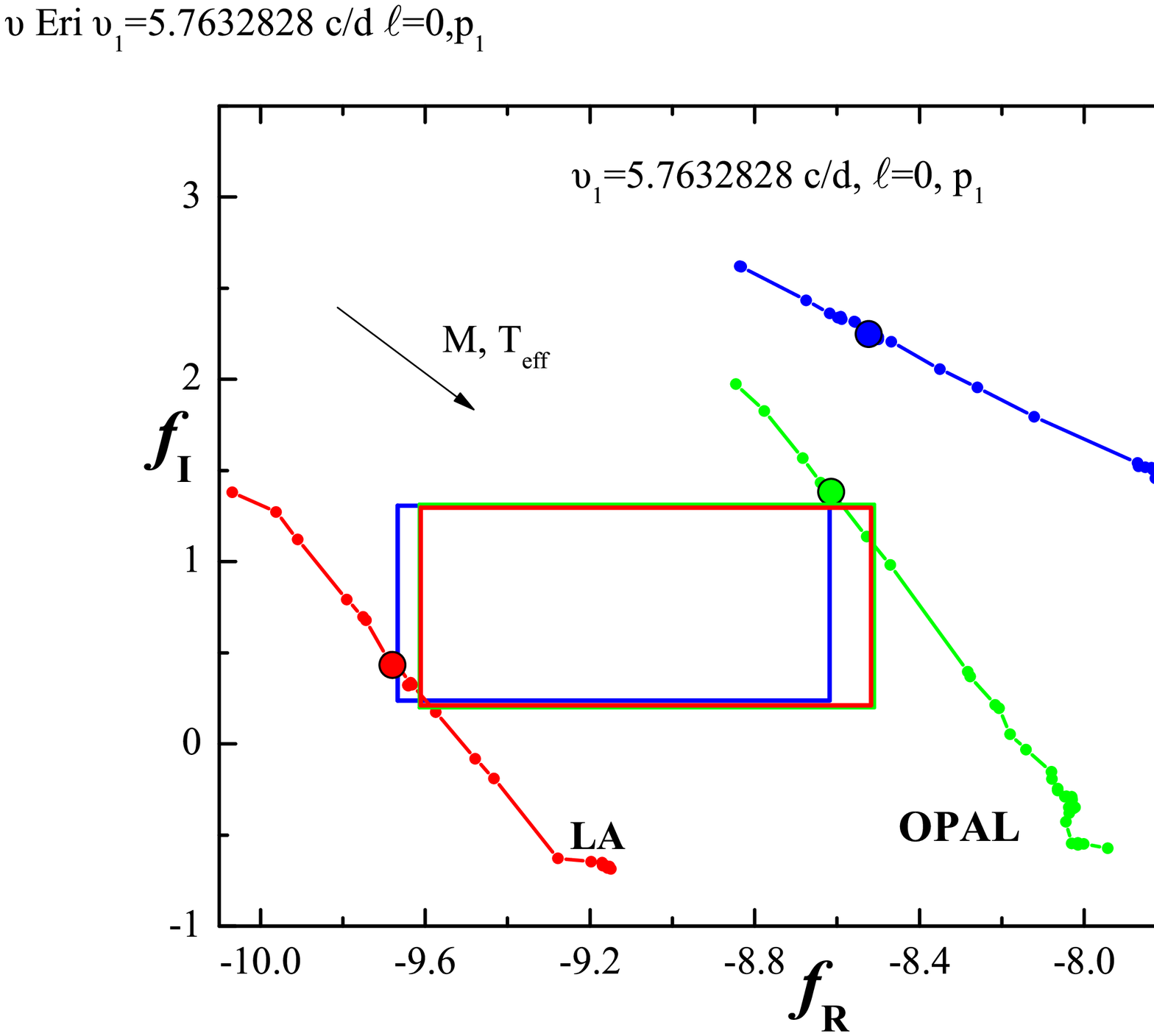}
 \includegraphics[clip,width=63mm]{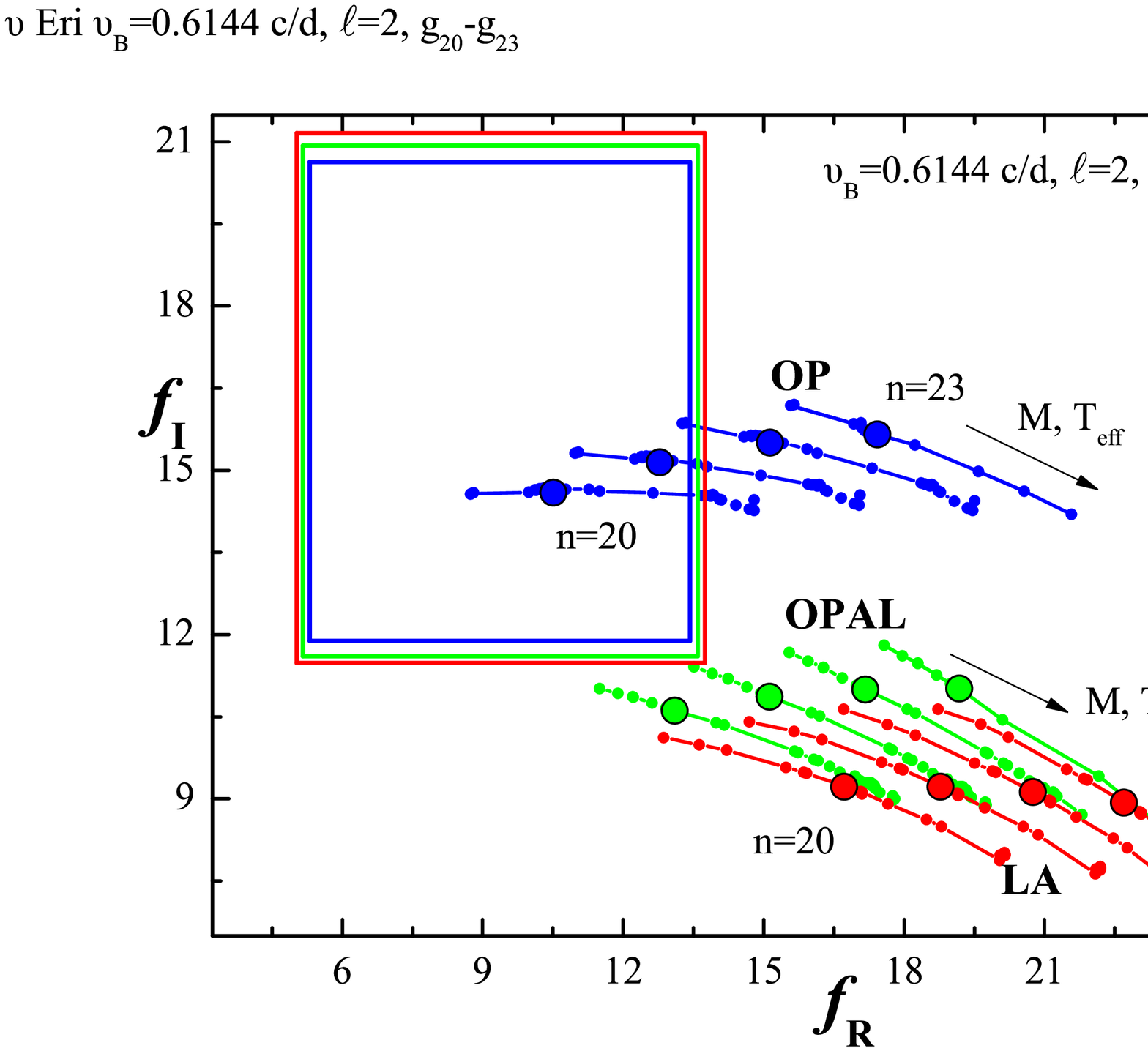}
 \vspace*{-0.2 cm}
 \caption{Comparison of the empirical (box) and theoretical (line) values of the $f$-parameter for the $\nu_1$ mode (left panel) and $\nu_{\rm B}$ mode (right panel).}
   \label{nuEri2}
\end{center}
\end{figure}

In Fig.\,\ref{nuEri2}, we show a comparison of the empirical and theoretical values of the non-adiabatic $f$-parameter for two modes: radial $\nu_1$ (left panel) and quadrupole $\nu_{\rm B}=0.6144$\,d$^{-1}$ (right panel) which is of the SPB type. We plotted the imaginary part of the $f$-parameter as a function of its real part. The boxes represent empirical values and the lines - theoretical.~The large dots mark models fitting also the $\nu_9$ frequency.

We see that the agreement is rather poor. In the case of the radial mode, we have some marginal agreement for the LA and OPAL models. For the $\nu_{\rm B}$ frequency we plotted a few modes with different radial orders (from $n=20$ to $n=23$). These models do not fit exactly the $\nu_{\rm B}$ frequency, but are very close to it. In this case, the OP models fit the empirical value of the $f$-parameter much better than the LA or OPAL models.

\begin{figure}[h]
 \vspace*{-.2 cm}
\begin{center}
 \includegraphics[clip,width=63mm,height=53mm]{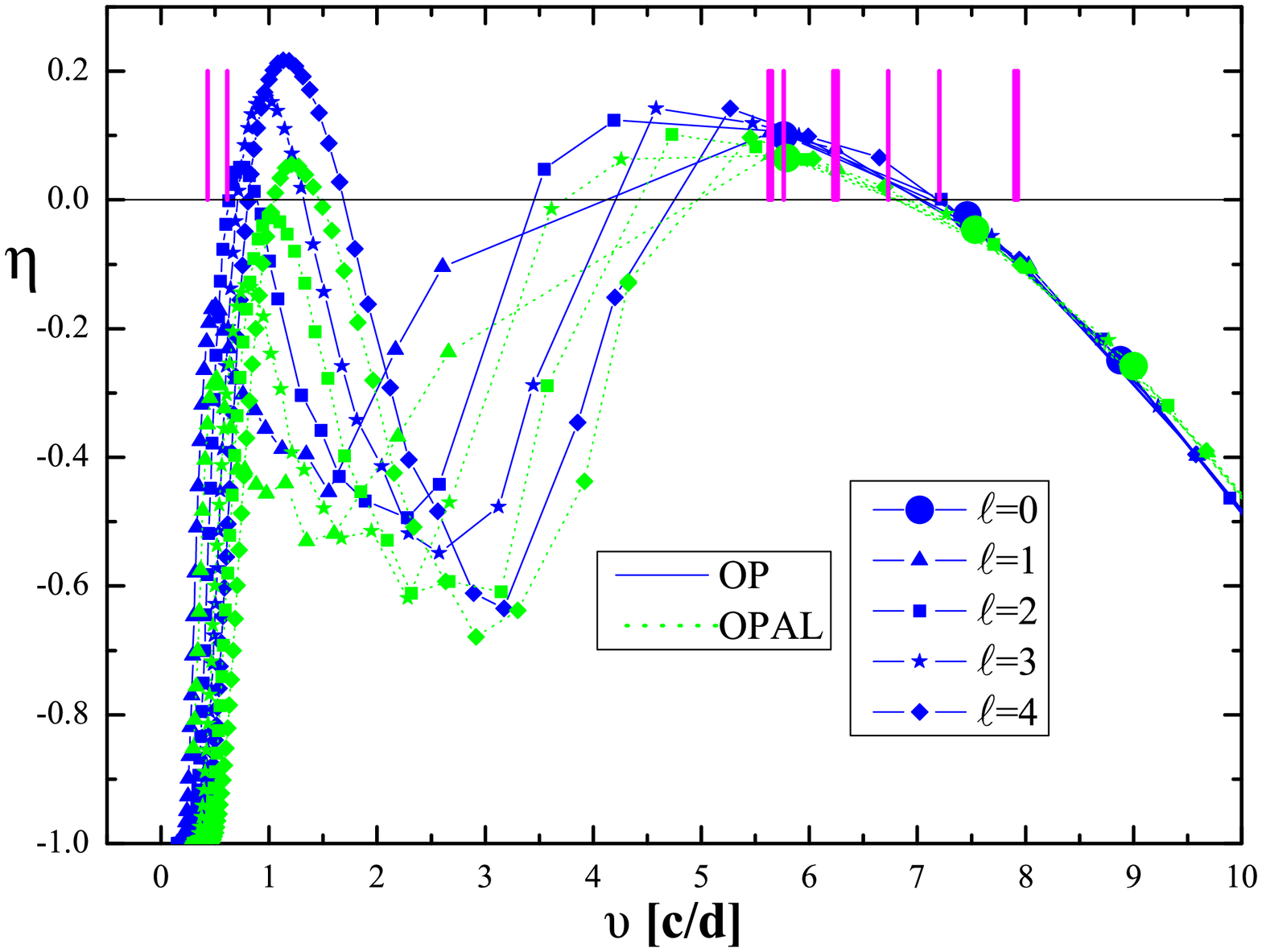}
 \includegraphics[clip,width=63mm,height=53mm]{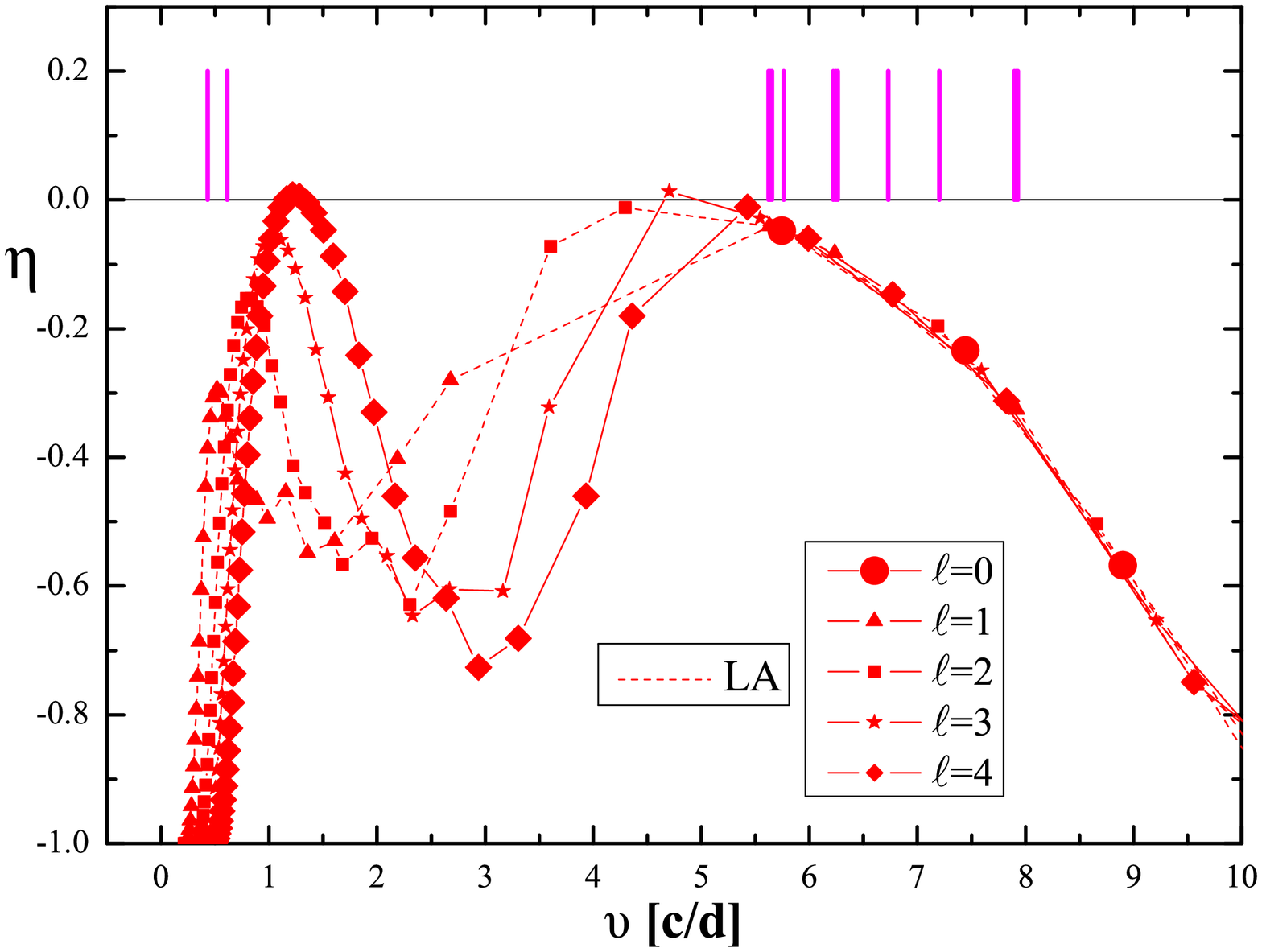}
 \vspace*{-0.2 cm}
 \caption{Instability parameter, $\eta$, as a function of the frequency for three seismic models of $\nu$ Eri calculated with the OP and OPAL data (left panel) and LA opacities (right panel).}
   \label{nuEri3}

\end{center}
  \vspace*{-0.2 cm}
\end{figure}

 \vspace*{-0.3 cm}
The low metallicity of the LA models as well as the low value of the opacity coefficient itself cause large problems with excitation of modes. In Fig.\,\ref{nuEri3} we plotted the instability parameter, $\eta$, as a function of frequency for three seismic models of $\nu$ Eri computed with the OP and OPAL opacities (left panel) and LA tables (right panel). The short vertical lines correspond to frequency spectrum of $\nu$ Eri. We can see that the LA model is almost entirely stable. The OP and OPAL models cannot excite the high frequency modes. Also the very low frequencies are stable, especially for OPAL opacities.

\section{Summary}
The B-type pulsators are very suitable for testing the opacity tables because a small difference in $\kappa$ results in quite large differences in seismic models. We could see that both frequencies and the $f$-parameter are sensitive to the opacities.

We found that, in case of $\theta$ Oph, the OPAL tables are the best. 12 Lac prefers OP data, while models of $\gamma$ Peg are rather similar with the OP and OPAL opacities. The LA data seems not to be good for $\gamma$ Peg. For $\nu$ Eri, the $f$-parameter of the radial mode prefers the LA or OPAL opacities, but the SPB-type mode favors instead the OP tables.
It seems that, in some parameter space, the OP opacities are better, in others - the OPAL data. The LA opacity table values are definitely too low; they are much smaller than OP or OPAL, especially in the region of the $Z$ bump, where the differences reach 9 to 10 \%.

Although, the presented results are not unambiguous, they show that further improvements and corrections in the opacity computations are needed.

\vspace*{0.5 cm}
\textbf{Acknowledgment.} Calculations have been carried out using resources provided by the Wroclaw Centre for Networking and Supercomputing (http://wcss.pl), grant No.~265.

\end{document}